\documentclass[aps,pra,reprint,superscriptaddress]{revtex4-1}

\usepackage{CJK}
\usepackage{graphicx,amssymb,bm,amsfonts,amsmath,graphics,color,epstopdf}
\usepackage[bookmarks=false,pdfstartview=FitH,colorlinks=true,citecolor=blue,linkcolor=blue]{hyperref}
\newcommand {\bscco}{Bi$_2$Sr$_2$CaCu$_2$O$_{8+\delta}$}
\newcommand {\uJcm}{$\mu$J/cm$^2$}

\begin{document}
\begin{CJK*}{GBK}{}

%\preprint{}

\title{Influence of Optically Quenched Superconductivity on Quasiparticle Relaxation Rates in \bscco}

\author{Christopher L.\ Smallwood}
\altaffiliation[Present address: ]{JILA, University of Colorado \& National Institute of Standards and Technology, Boulder, CO 80309, USA}%Lines break automatically or can be forced with \\
\email[Email: ]{chris.smallwood@colorado.edu}
\affiliation{Materials Sciences Division, Lawrence Berkeley National Laboratory, Berkeley, California 94720, USA}
\affiliation{Department of Physics, University of California, Berkeley, California 94720, USA}
\author{Wentao Zhang}
\affiliation{Materials Sciences Division, Lawrence Berkeley National Laboratory, Berkeley, California 94720, USA}
\affiliation{Department of Physics, University of California, Berkeley, California 94720, USA}
\author{Tristan L.\ Miller}
\affiliation{Materials Sciences Division, Lawrence Berkeley National Laboratory, Berkeley, California 94720, USA}
\affiliation{Department of Physics, University of California, Berkeley, California 94720, USA}
\author{Gregory Affeldt}
\affiliation{Materials Sciences Division, Lawrence Berkeley National Laboratory, Berkeley, California 94720, USA}
\affiliation{Department of Physics, University of California, Berkeley, California 94720, USA}
\author{Koshi Kurashima}
\affiliation{Department of Applied Physics, Tohoku University, Sendai 980-8579, Japan}
\author{Chris Jozwiak}
\affiliation{Advanced Light Source, Lawrence Berkeley National Laboratory, Berkeley, California 94720, USA}
\author{Takashi Noji}
\affiliation{Department of Applied Physics, Tohoku University, Sendai 980-8579, Japan}
\author{Yoji Koike}
\affiliation{Department of Applied Physics, Tohoku University, Sendai 980-8579, Japan}
\author{Hiroshi Eisaki}
\affiliation{Electronics and Photonics Research Institute, National Institute of Advanced Industrial Science and Technology, Tsukuba, Ibaraki 305-8568, Japan}
\author{Dung-Hai Lee}
\affiliation{Materials Sciences Division, Lawrence Berkeley National Laboratory, Berkeley, California 94720, USA}
\affiliation{Department of Physics, University of California, Berkeley, California 94720, USA}
\author{Robert A.\ Kaindl}
\affiliation{Materials Sciences Division, Lawrence Berkeley National Laboratory, Berkeley, California 94720, USA}
\author{Alessandra Lanzara}
\email[Email: ]{alanzara@lbl.gov}
\affiliation{Materials Sciences Division, Lawrence Berkeley National Laboratory, Berkeley, California 94720, USA}
\affiliation{Department of Physics, University of California, Berkeley, California 94720, USA}
\date {\today}

\begin{abstract}
We use time- and angle-resolved photoemission to measure quasiparticle relaxation dynamics across a laser-induced superconducting phase transition in \bscco.
Whereas low-fluence measurements reveal picosecond dynamics, sharp femtosecond dynamics emerge at higher fluence.
Analyses of data as a function of energy, momentum, and doping indicate that the closure of the near-nodal gap and disruption of macroscopic coherence are primary mechanisms driving this onset.
The results demonstrate the important influence of transient electronic structure on relaxation dynamics, which is relevant for developing an understanding of nonequilibrium phase transitions.
\end{abstract}

% 74.40.Gh  Nonequilibrium superconductivity
% 74.25.Jb  Electronic structure (photoemission, etc.) of superconductors
% 78.47.J-  Ultrafast spectroscopy
% 74.72.Gh  Cuprate superconductors, hole doped 
%\keywords{}

\maketitle
\end{CJK*} 

% If in two-column mode, this environment will change to single-column
% format so that long equations can be displayed. Use
%\begin{widetext}
%\end{widetext}

% figures should be put into the text as floats.
% Use the graphics or graphicx packages (distributed with LaTeX2e)
% and the \includegraphics macro defined in those packages.
% See the LaTeX Graphics Companion by Michel Goosens, Sebastian Rahtz,
% entire page. There is no need to do explicit centering.

%Paragraph 1
In the study of complex many-body interactions, a fundamental topic is the behavior of materials at the boundaries of ordered phases, which can be accessed, for example, by monitoring material properties as a function of temperature, magnetic field, or chemical doping. Recently, there has been great interest in studying phase transitions not just in steady state, but also out of equilibrium~\cite{Orenstein12,Zhang14a}. It has been shown, for example, that ultrafast optical pulses can probe and in cases even manipulate electronic order in materials ranging from high-temperature superconductors~\cite{Demsar99,Kaindl00,Gedik04,Fausti11,Rettig12,Smallwood12,Torchinsky13,Hu14}, to topological insulators~\cite{Sobota12,Wang13}, to charge-density-wave materials~\cite{Schmitt08,Rohwer11,Stojchevska14}. Although increasingly powerful theoretical techniques are being developed~\cite{Howell04,Sentef13,Kemper14,Sentef15,Sentef15a}, our understanding of these phenomena remains lacking. 

%Paragraph 2
In high-temperature superconductors, particularly in the cuprates, an important related question is how quasiparticle interactions are influenced by electronic structure when the system transitions across the phase boundary between the superconducting and pseudogap regimes. An intriguing recent approach for insight into this physics has been the examination of quasiparticle relaxation using pump--probe spectroscopy at low temperature but high pump fluence, where the superconducting condensate is fully vaporized into quasiparticles and must dynamically re-emerge. This regime has been explored using time-resolved reflectivity and transmissivity~\cite{Demsar99,Liu08a,Kusar08,Coslovich11}, and studies have reported distinct femtosecond and picosecond relaxation time scales that were often assigned to the condensate and pseudogap phases. However, such all-optical probes lack momentum resolution, and have no direct access to the dynamics of the electronic band structure. Conversely, time- and angle-resolved photoemission spectroscopy (time-resolved ARPES)~\cite{Carpene09,Kirchmann10,Faure12,Smallwood12a} is ideally suited for probing both momentum-dependent quasiparticle~\cite{Nessler98,Perfetti07,Graf11,Cortes11,Smallwood12,Zhang13,Rameau14,Yang15,Piovera15} and gap~\cite{Smallwood12,Smallwood14,Zhang14} dynamics in cuprates, yet low-temperature studies using fluences high enough to destroy superconductivity~\cite{Cortes11,Graf11,Zhang13,Smallwood14,Zhang14,Yang15,Piovera15} have to date provided no comparisons of the gap and quasiparticle population.

%Paragraph 3
In the present study, we use time-resolved ARPES to probe the critical-fluence regime of the high-temperature superconductor \bscco\ (Bi2212) in detail, providing a characterization of quasiparticle relaxation rates as a simultaneous function of crystal momentum and energy in cuprates. Fluences inducing a complete closure of the near-nodal gap result in a distinctive two-component relaxation signature, with well-defined femtosecond and picosecond recovery scales that occur throughout a large range of $k$-space. Correlations between the onset of these two components and gap dynamics establish a strong connection between picosecond relaxation dynamics and superconductivity. However, the detailed momentum and doping dependence of the faster, femtosecond component contradicts an assignment to pseudogap-phase dynamics. Instead, the dynamics can be explained by release of kinetic restrictions within a ``normal-state'' phase obtained by a full quenching of the condensate. The results demonstrate the important influence of transient electronic structure on relaxation dynamics, which is relevant for developing an understanding of nonequilibrium phase transitions.

%Paragraph 4
The experimental apparatus is as described in Ref.~\cite{Smallwood12a}, and uses 836-nm ($h\nu=1.48$ eV) pump pulses and 209-nm ($h\nu=5.93$ eV) probe pulses. Energy, momentum, and time resolutions are 23 meV, 0.003 \AA$^{-1}$, and 300 fs. We have corrected for detector nonlinearity~\cite{Smallwood12a}, as well as for a pump-induced time-dependent  but uniform energy shift in the electronic spectrum ($\leq 4$ meV)~\cite{Smallwood14,Miller15}.

%%%%%%%%%%%%%%%%%%%%%%%%%%%%%%%%%
%Fig1
%%%%%%%%%%%%%%%%%%%%%%%%%%%%%%%%%
\begin{figure}[tb]\centering\includegraphics[width=3.375in]{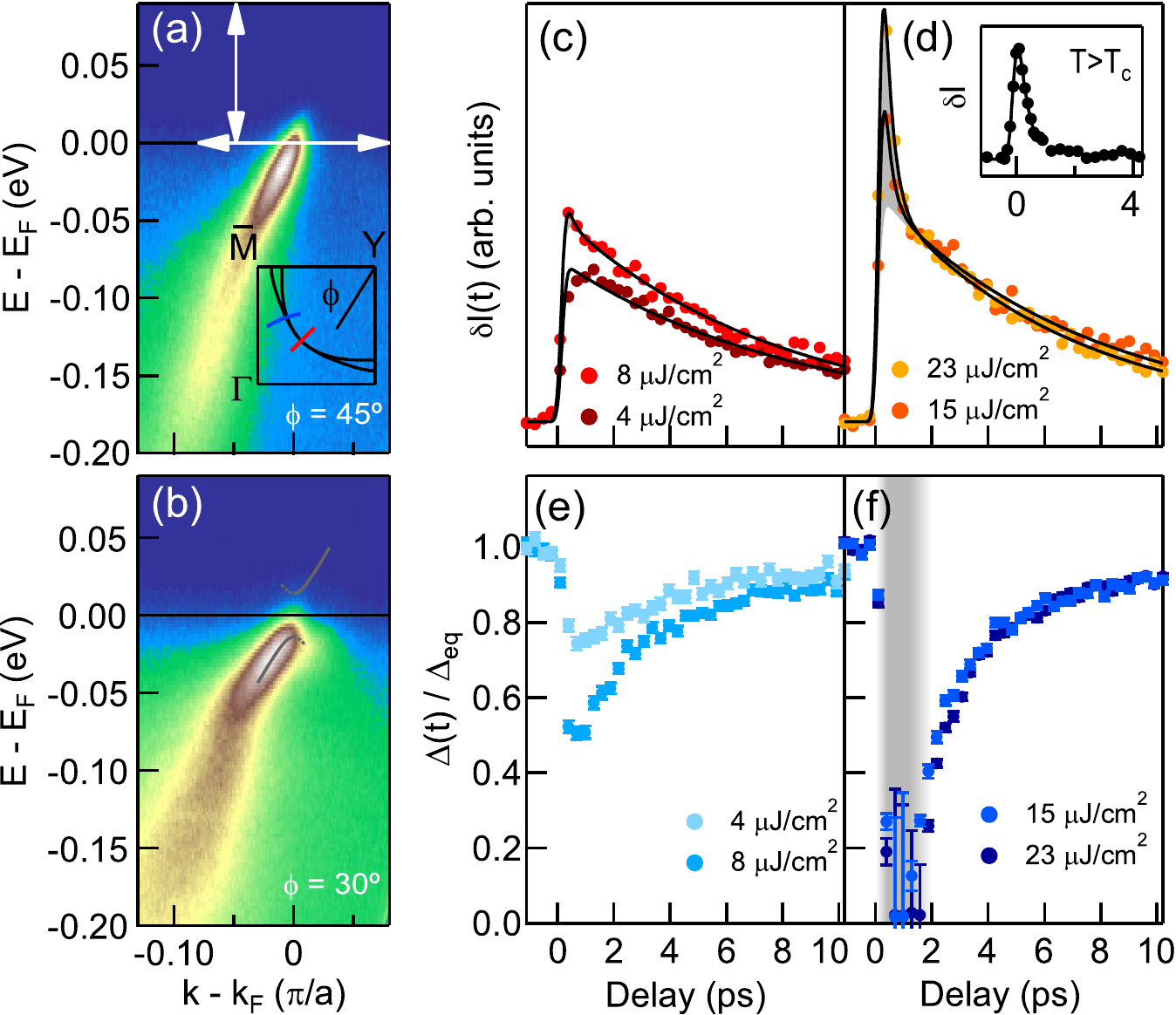}
\caption{\label{fig1}Evolution of the quasiparticle population and band gap in superconducting Bi2212 ($T_c=91$ K) following an ultrafast infrared pump pulse.
{\bf(a)--(b)} Equilibrium ARPES dispersions ($t=-1.1$ ps) for cuts at $\phi=45^\circ$ and $\phi=30^\circ$ (see panel (a) schematic). 
{\bf(c)--(d)} Nodal ($\phi=45^\circ$) quasiparticle population $\delta I(t)$, from integrating ARPES intensity between the white double arrows in (a).
Main panels show $T = 20$ K. Inset shows data at 23 \uJcm\ for $T=100$ K ($T>T_c$)\@.
{\bf(e)--(f)} Normalized superconducting gap $\Delta(t)$ from the cut at $\phi=30^\circ$, extracted by fitting symmetrized EDCs at $k_F$ to a broadened BCS line shape~\cite{Smallwood14}.
}
\end{figure}

%Paragraph 5
Figure~\ref{fig1} shows a fluence-dependent analysis of a nearly optimally doped sample ($T_c=91$ K), where two cuts through $k$-space are depicted, at $\phi=45^\circ$ and $\phi=30^\circ$ ($\phi$ is defined from the $Y$ point relative to $Y$--$\bar{M}$ as in the panel (a) inset). The cut at $\phi=45^\circ$ intersects a $d$-wave gap node, where quasiparticle dynamics can be measured independently of gap dynamics, and can be characterized with a high signal-to-noise ratio (Figs.~\ref{fig1}(c) and \ref{fig1}(d)) by integrating spectral intensity change across a large window in energy and momentum (white double arrows in Fig.~\ref{fig1}(a)). Concomitant gap dynamics (Figs.~\ref{fig1}(e) and \ref{fig1}(f)) are extracted using energy distribution curve (EDC) symmetrization~\cite{Norman98a,Smallwood12,Smallwood14} from the cut at $\phi=30^\circ$, which has an equilibrium gap of 14 meV (Fig.~\ref{fig1}(b)). Consistent with previous measurements~\cite{Smallwood12}, at lower fluence the gap magnitude is suppressed but always finite, and the population of nonequilibrium quasiparticles decays with picosecond dynamics. As shown in Figs.~\ref{fig1}(d) and \ref{fig1}(f), however, beyond a critical fluence $F_c$ (defined as the fluence necessary to close the near-nodal gap)~\cite{Smallwood14,Zhang14}, a sharp femtosecond relaxation signature emerges in the quasiparticle population response. The signature is also coupled to the near-nodal gap temporally, appearing most prominently when the gap is closed ($t<1.5$ ps, see shaded gray regions in Figs.~\ref{fig1}(d) and \ref{fig1}(f)). We note that, although two-component quasiparticle recovery dynamics have been previously observed in cuprate studies using time-resolved ARPES~\cite{Perfetti07,Graf11,Cortes11} and time-resolved reflectivity and transmissivity~\cite{Demsar99,Kaindl00,Liu08a,Kusar08,Coslovich11,DalConte12,Coslovich13,Cilento14}, this connection between two-component behavior and the superconducting gap has not been reported.

\begin{figure}[tb]\centering\includegraphics[width=3.25in]{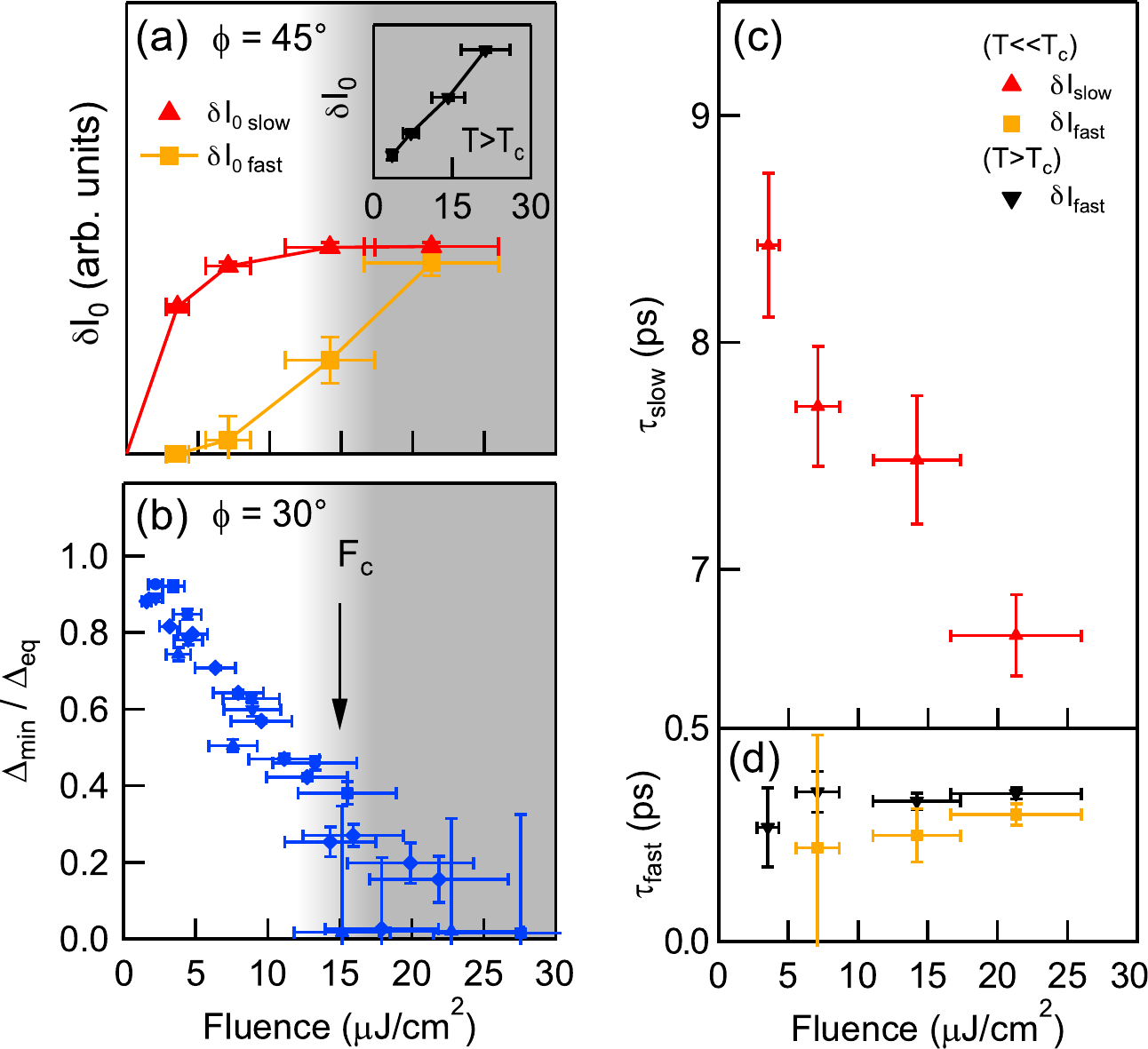}
\caption{\label{fig2}
{\bf(a)} Amplitudes $\delta I_{0 \, slow}$ and $\delta I_{0fast}$, from fits to the data in Figs.~\ref{fig1}(c) and \ref{fig1}(d) as described in the text~\cite{sm}.
{\bf(b)} Minimal gap from Figs.~\ref{fig1}(e) and \ref{fig1}(f) versus fluence.
{\bf(c)--(d)} Time constants $\tau_{slow}$ and $\tau_{fast}$, from fits to the data in Figs.~\ref{fig1}(c) and \ref{fig1}(d) as described in the text~\cite{sm}.
}
\end{figure}

%Paragraph 6
Figure~\ref{fig2} shows a more detailed analysis of quasiparticle decay, where fast and slow component amplitudes and time constants are extracted from Fig.~\ref{fig1} by fitting the integrated ARPES intensity with a bi-exponential decay
\begin{equation}
\delta I(t) = \Theta(t) \left[  C_{fast} e^{-t/\tau_{fast}} + C_{slow} e^{-t/\tau_{slow}} \right]
\label{biexp}
\end{equation} 
after convolution with a Gaussian of 300-fs duration to incorporate time resolution. In the above, $t \equiv \text{Delay}-t_0$, and $\tau_{fast}$ and $\tau_{slow}$ are the time constants of the respective fast and slow components of the relaxation, while $C_{fast}$ and $C_{slow}$ are their bare amplitudes. After convolution with the resolution function, we extract the effective amplitudes $\delta I_{0 fast}$ and $\delta I_{0 \, slow}$, which directly relate to the data~\cite{sm}.
For the 4 \uJcm\ measurement we constrained $C_{fast}=0$ because a fast component was not apparent. We also measured the above-$T_c$ response, setting $C_{slow}=0$ and constraining $t < 0.6$ ps because a slow component was not apparent.

%%%%%%%%%%%%%%%%%%%%%%%%%%%%%%%%%
%Fig2
%%%%%%%%%%%%%%%%%%%%%%%%%%%%%%%%%

%Paragraph 7
Overall, the picosecond quasiparticle recovery component exhibits many trends that establish a close connection to superconductivity. For example, the picosecond component only appears in the low-temperature data. As shown in Fig.~\ref{fig2}(c), when present, $\tau_{slow}$ decreases with increasing fluence~\cite{Smallwood12}, as expected of quasiparticles reentering a superconducting condensate following rules of second-order kinetics~\cite{Gedik04,Kaindl05,Smallwood12}. The amplitude $\delta I_{0 \, slow}$ also increases rapidly with fluence in the low-fluence limit, yet appears to saturate above $F_c$ (Fig.~\ref{fig2}(a)), which is consistent with a density of quasiparticles decaying into the superconducting state in a manner fundamentally limited by the equilibrium superfluid density. By contrast, the femtosecond recovery component exhibits very different behavior. At low temperature, $\delta I_{0 fast}$ is suppressed or absent at low fluence and only becomes substantial for $F>F_c$. At high temperature (above $T_c$), femtosecond dynamics dominate. At all temperatures, $\tau_{fast}$ exhibits no fluence dependence or may even increase with fluence (Fig.~\ref{fig2}(d)).

%%%%%%%%%%%%%%%%%%%%%%%%%%%%%%%%%
%Fig3
%%%%%%%%%%%%%%%%%%%%%%%%%%%%%%%%%
\begin{figure}[tb]\centering\includegraphics[width=3.2in]{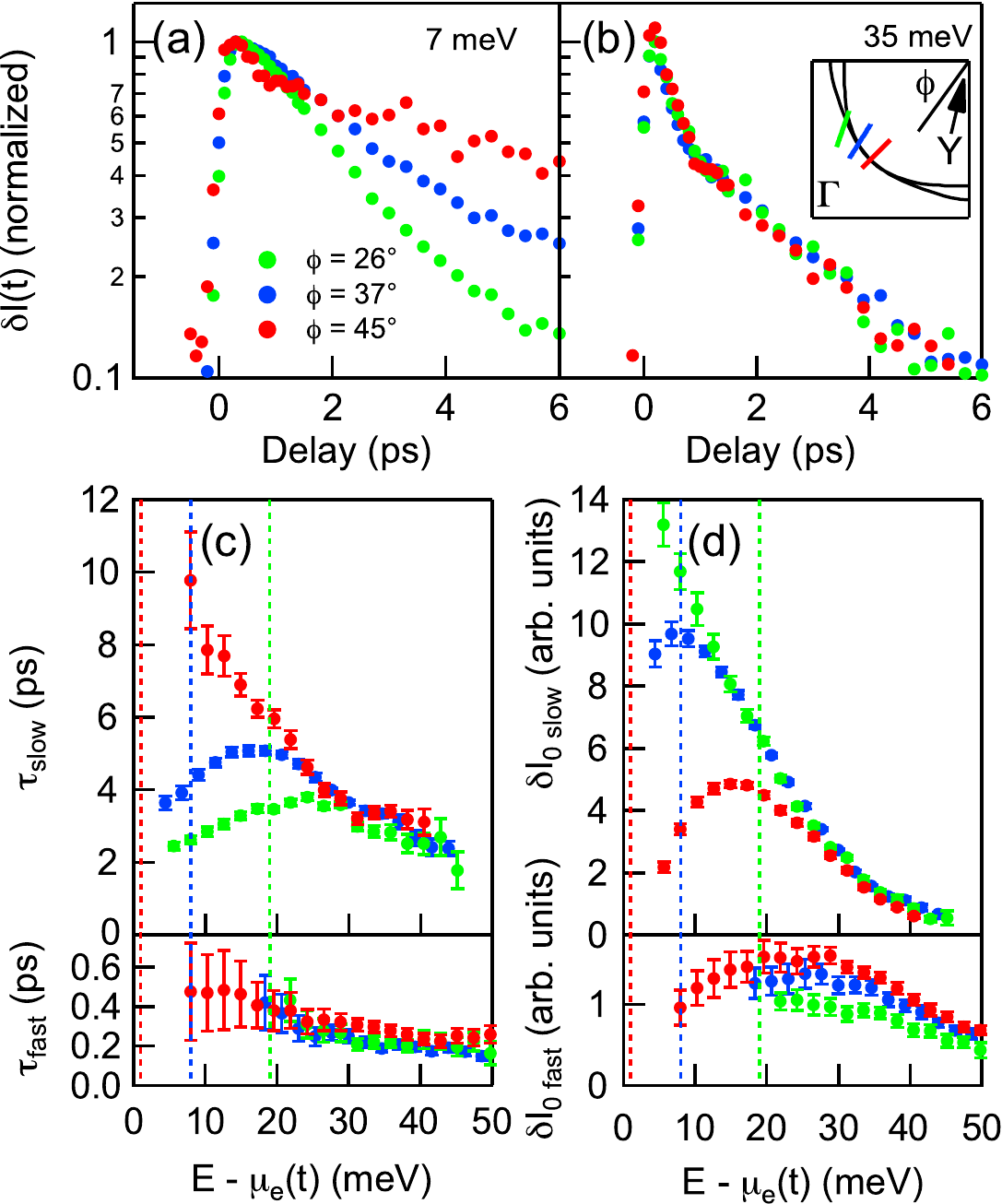}
\caption{\label{qpmom}Energy and momentum dependence of quasiparticle recombination dynamics in optimally doped Bi2212 (25 \uJcm\ pump fluence, $T = 20$ K, $T_c=91$ K).
{\bf(a)--(b)} Energy-resolved ARPES intensity change $\delta I(t)$, momentum-integrated between $|k-k_F|<0.08$ $\pi/a$ for nodal, off-nodal, and far-off-nodal cuts, at two representative energies above $E_F$. 
{\bf(c)--(d)} Energy-resolved quasiparticle relaxation time constants and amplitudes.
The red, blue, and green vertical dashed lines mark the gap edge at $\phi=45^\circ$, $\phi=37^\circ$, and $\phi=26^\circ$.
Intensities in (d) are obtained by scaling momentum cuts so that $\int_{-0.3 \text{eV}}^{0.1 \text{eV}} I(k,\omega) \, d\omega$ averages to 1 for $k$ such that $-0.075 \pi/a<k-k_F < -0.03 \pi/a$ when $t<0$.
}
\end{figure}

%Paragraph 8
Figure~\ref{qpmom} shows quasiparticle relaxation dynamics, resolved in both energy and momentum, for momentum cuts at $\phi=26^\circ$ (outside the pseudogap-state Fermi arc), $\phi=37^\circ$ (inside the pseudogap-state Fermi arc), and $\phi=45^\circ$ (the node), responding to a pump fluence of 25 \uJcm\ (above $F_c$). Such an analysis takes advantage of the full power of time-resolved ARPES, and two-component relaxation dynamics notably appear throughout a large portion of the Brillouin zone. As clear from Figs.~\ref{qpmom}(a) and \ref{qpmom}(c), quasiparticle relaxation after the first picosecond occurs significantly faster away from the node than at the node, in agreement with previous energy-integrated measurements at lower fluence~\cite{Smallwood12}. These momentum variations in $\tau_{slow}$ become larger with decreasing energy (Fig.~\ref{qpmom}(b)), and establish further connections to superconductivity, as low-energy quasiparticles necessarily interact more strongly with the many-body ground state than do their higher-energy counterparts~\cite{Tinkham}.

%%%%%%%%%%%%%%%%%%%%%%%%%%%%%%%%%
%Fig4
%%%%%%%%%%%%%%%%%%%%%%%%%%%%%%%%%
\begin{figure*}[!htbp]\centering\includegraphics[width=6.5in]{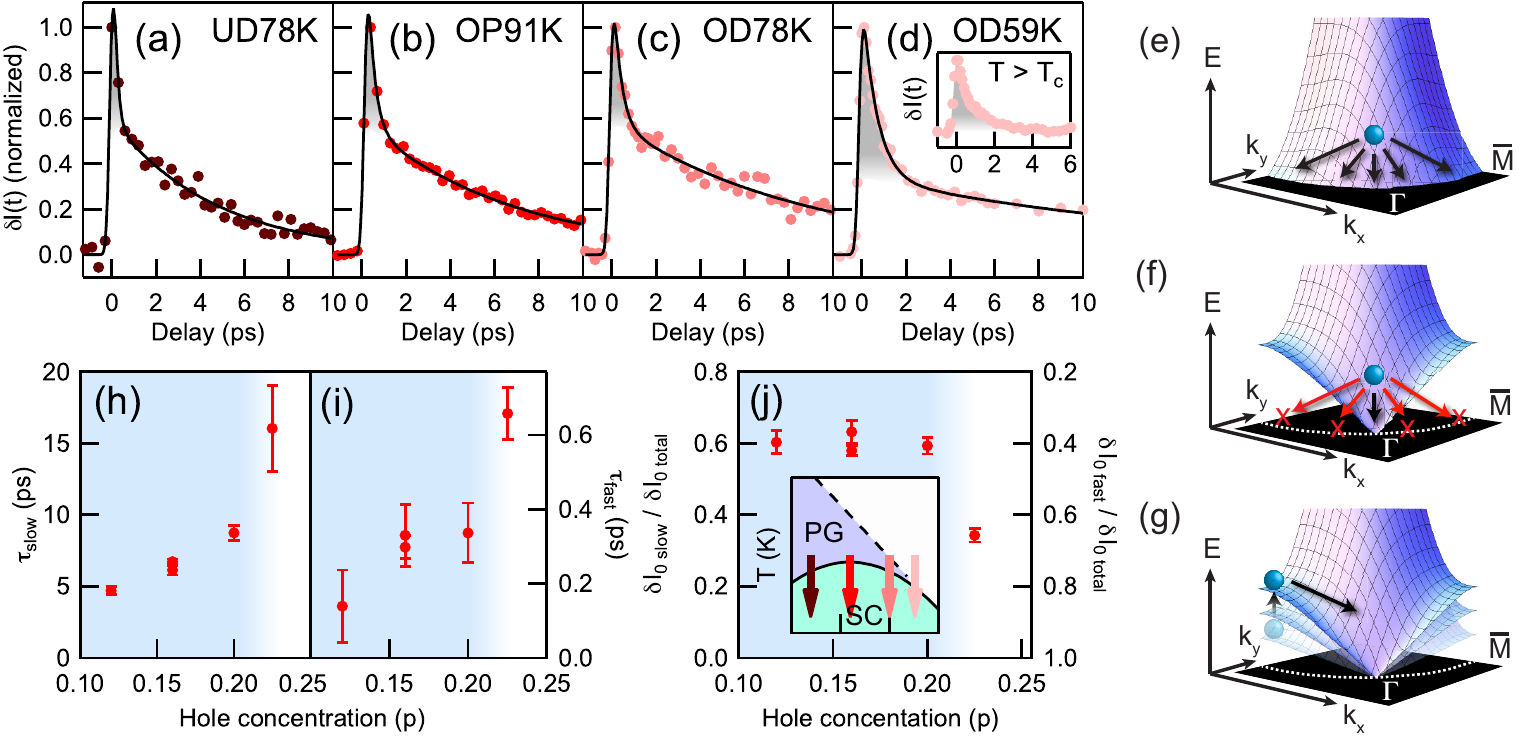}
\caption{\label{doping}Doping dependence of quasiparticle decay for equilibrium temperature $T = 20$ K (unless otherwise noted) and fluence $F = 23$ \uJcm. 
{\bf(a)--(d)} Nodal quasiparticle response curves, extracted as in Fig.~\ref{fig1}. The fits are bi-exponential decay functions (Eq.~\ref{biexp}).
Panel (d) inset shows the OD59K nodal quasiparticle response at $T=70$ K.
{\bf(e)--(g)} Cartoon illustrations of the impact on quasiparticle scattering of opening a $d$-wave gap.
{\bf(h)--(i)} Time constants from the fits in (a)--(d)~\cite{sm}.
{\bf(j)} Amplitude component ratios $\delta I_{0 \, slow}/(\delta I_{0 \, slow}+\delta I_{0 \, fast})$ from the fits in (a)--(d). 
Inset is the equilibrium Bi2212 phase diagram, showing the relationship between doping, the pseudogap (PG), and superconductivity (SC)~\cite{Chatterjee11,Vishik12}.
}
\end{figure*}

%Paragraph 9
Though it is tempting, based on Figs.~\ref{fig1} and \ref{fig2}, to associate the fast component with a competing pseudogap order, Fig.~\ref{qpmom} shows that the momentum- and energy-dependent trends in $\delta I_{0 fast}$ are directly at odds with the fact that the pseudogap becomes most prominent toward the Brillouin zone face. At optimal doping, the equilibrium pseudogap is absent or undetectable near $\phi=45^\circ$, and becomes increasingly pronounced toward $\phi=0^\circ$, starting at $\phi \approx 30^\circ$~\cite{Lee07,Vishik12}. As shown in Fig.~\ref{qpmom}(a), by contrast, at low energy the fast component completely disappears in the off-nodal data. In part, this reflects the fact that the low-energy off-nodal quasiparticle relaxation signature is overshadowed by the dynamics of the nonequilibrium superconducting gap. However, even above the gap edge,  $\tau_{fast}$ exhibits no discernible momentum dependence within experimental error (Figs.~\ref{qpmom}(b) and \ref{qpmom}(c)), and energy-dependent amplitudes $\delta I_{0 fast}$ do not appear to increase near the Brillouin zone face (Fig.~\ref{qpmom}(d)). Rather, they may even tend to decrease, although ARPES matrix elements make it somewhat difficult to make these comparisons quantitative.

%Paragraph 10
Stronger evidence differentiating the fast component from the physics of the pseudogap occurs in the doping dependence of quasiparticle relaxation. Figure~\ref{doping} shows nodal quasiparticle relaxation dynamics at comparable excitation densities ($\approx 24$ \uJcm) for multiple dopings of Bi2212, corresponding to critical temperatures $T_c=78$ K (underdoped, UD78K), $T_c=91$ K (nearly optimally doped, OP91K), $T_c=78$ K (overdoped, OD78K), and $T_c=59$ K (very overdoped, OD59K)\@. For the first three dopings, both equilibrium ARPES~\cite{Chatterjee11,Vishik12} and time-resolved ARPES~\cite{Zhang13} measurements have previously identified distinct transition temperatures $T_c$ and $T^*$ for the onset of superconductivity and the pseudogap, with $T^*$ occurring at 200 K for the UD78K sample, 150 K for the OP91K sample, and 97 K for the OD78K sample. The OD59K sample has no equilibrium pseudogap~\cite{Chatterjee11,Vishik12}. In spite of these differences, two-component recovery dynamics prominently appear in the time-resolved nodal quasiparticle relaxation dynamics for all four dopings (Figs.~\ref{doping}(a)--\ref{doping}(d)), and quasiparticle relaxation above $T_c$ in the OD59K sample resembles the low-temperature femtosecond component for the OD59K sample (Fig.~\ref{doping}(d)) similarly to the way that quasiparticle relaxation above $T_c$, but below $T^*$, in the OP91K sample resembles the OP91K low-temperature femtosecond relaxation component (Fig.~\ref{fig1}(d))~\cite{sm}.

%Paragraph 11
Taking the momentum and doping dependence of quasiparticle relaxation rates into consideration jointly, we argue that the primary explanation for the onset of two-component dynamics is that quasiparticle relaxation dynamics are directly affected by changes that occur in the quasiparticle spectrum as the system relaxes back toward equilibrium. More specifically, the opening of a gap dramatically reduces the amount of available phase space near the Fermi level. A quasiparticle in metallic Bi2212 can relax by many channels (Fig.~\ref{doping}(e)), with a final-state phase-space density roughly proportional to the product of the Fermi surface area and the quasiparticle energy (assuming a uniform Fermi velocity). In the presence of a $d$-wave gap, however, scattering rates for lower-energy states are sharply curtailed (Fig.~\ref{doping}(f)). For a quasiparticle residing along the nodal direction at 15 meV in optimally doped Bi2212 ($\Delta_0 \approx 35$ meV), the gap imposes an almost 80\% phase-space reduction for relaxation via scattering (particle-conserving) interactions~\footnote{The $d$-wave gap is estimated near the node as $\Delta_k(\phi) \approx v_{gap} (\phi-45)\pi/180$, where $\phi$ is the Fermi surface angle defined in the main text. Assuming a cylindrical Fermi surface, the fraction of final-state phase space (states between $0$ and $E$) relative to ungapped final-state phase space is then given by $E / v_{gap}$, where $E$ is the quasiparticle initial-state energy and $v_{gap} \approx 66$ meV/radian is estimated from the literature~\cite{Vishik10}.}.
Beyond this, the gap opens in the midst of the quasiparticle relaxation process, which means that quasiparticles are steadily lifted from lower energies back to higher energies on a picosecond timescale (Fig.~\ref{doping}(g)), and the overall effect counterbalances quasiparticle relaxation at fixed energy. This explains, for example, why the energy of the maximum $\tau_{slow}$ increases with decreasing $\phi$ in Fig.~\ref{qpmom}(c). While the effect occurs most dramatically at the Brillouin zone face, nodal and near-nodal states might also be impacted because of the resultant conversion of boson-absorption scattering channels into boson-emission channels.
We note that momentum conservation requirements are ignored in these first two arguments,
which may be reasonable given the large inhomogeneities that are known to exist in Bi2212 on nanometer length scales~\cite{Pan01}.
Finally, there is a fundamental change---due to coherence factors---in electron-boson coupling matrix elements in the presence as opposed to the absence of superconductivity, which may affect quasiparticle recombination rates in important ways~\cite{Tinkham,Hanaguri09}. For example, if the dominant channel for quasiparticle recombination is between $k$-space regions where the gap parameter $\Delta_k$ has like signs, and the dominant scattering interaction is odd under time reversal, then recombination will be reduced in the superconducting state as compared to the normal state.

%Paragraph 12
Curiously, Figs.~\ref{doping}(h) and \ref{doping}(i) reveal that time constants for {\it both} the fast and slow components decrease sharply with decreasing hole concentration, and the amplitude ratio $\delta I_{0 \, slow}/\delta I_{0 \,\text{total}}$ is markedly different for the OD59K sample than it is for the other three dopings (Fig.~\ref{doping}(j)). Further studies with better time resolution are needed to investigate these effects more fully.

%Paragraph 13
In summary, we have shown that kinetically released quasiparticle relaxation dynamics and the opening of the superconducting gap play a large role in generating two-component relaxation dynamics, independently of the role that the pseudogap might play. This result will be important for any optically-induced phase transition involving a gap, but is particularly relevant for cuprates given the findings of many all-optical studies, which also report a femtosecond relaxation component, yet have associated it in many cases as a direct manifestation of the pseudogap~\cite{Demsar99,Kaindl00,Liu08a,Kusar08,Coslovich11,DalConte12,Coslovich13,Cilento14}. We note that all-optical studies are typically sensitive to energies far higher than those probed in the current work. Hence, the two techniques may be accessing different dynamics if, for example, the influence of the pseudogap extends to higher energies than does the superconducting gap, or if the low-energy signatures of the pseudogap are more localized in momentum than higher-energy signatures. Time-resolved ARPES studies at higher fluences or closer to the first Brillouin zone boundary may be helpful in resolving this question unambiguously.

\begin{acknowledgments}
We thank J.\ Orenstein, J.~P.\ Hinton, A.~Vishwanath, C.~M.\ Varma, A.~F.\ Kemper, D.\ Mihailovic, and C.-Y.\ Lin for useful discussions.
This work was supported as part of the Ultrafast Materials Program at Lawrence Berkeley National Laboratory, funded by the U.S.\ Department of Energy, Office of Science, Office of Basic Energy Sciences, Materials Sciences and Engineering Division, under Contract No.\ DE-AC02-05CH11231.
\end{acknowledgments}

%\bibliography{/Users/Chris2/Documents/TexDocuments/clsbib}

\end{document}